# A generic method for modeling the behavior of

# anisotropic metallic materials : application to

# recrystallized zirconium alloys

S. Leclercq[1,*], G. Rousselier[2], G. Cailletaud[2]

[1]EDF R&D, Département Matériaux et Mécanique des Composants

Les Renardières, 77818 Moret sur Loing Cedex, France

[2]Centre des Matériaux, ENSMP, UMR CNRS 7633, BP 87, 91003 Evry, France

Running title : Modeling of the behavior of anisotropic metallic materials

[*] Corresponding author. Tel.:  33 1 60 73 64 92; fax: 33 1 60 73 65 59
*E-mail address*: sylvain.leclercq@edf.fr

**Abstract**

A simplified polycrystalline model (the so-called RL model) is proposed to simulate the anisotropic viscoplastic behavior of metallic materials. A generic method is presented that makes it possible to build a simplified anisotropic material texture, based on the principal features of the pole figures. The method is applied to a recrystallised zirconium alloy, used as clad material in the fuel rods of nuclear power plants. An important data base consisting in mechanical tests performed on Zircaloy tubes is collected. Only a small number of tests (pure tension, pure shear) are used to identify the material parameters, and the texture parameters. It is shown that 6 crystallographic orientations (6 "grains") are sufficient to describe the large anisotropy of such hcp alloy. The identified crystallographic orientations match the experimental pole figures of the material, not used in the identification procedure. Special attention is paid to the predictive ability of the model, i.e. its ability to simulate correctly experimental tests not belonging to the identification data base. These predictive results are good, thanks to an identification procedure that enables to consider the contribution of each slip system in each crystallographic orientation.

*Keywords:* anisotropic material; constitutive behaviour; crystal plasticity; polycrystalline material; Zirconium alloys.

## 1. Introduction

Since the pioneering work of Hill (1950), several efforts have been made to improve the description of initial plastic anisotropy according to macroscopic constitutive equations (see for instance, Gotoh, 1977; Barlat et al., 1991; Karafillis and Boyce, 1993; Bron and Besson, 2004). Nevertheless, these macroscopic models generally fail to accurately represent non proportional and cyclic loadings (Cailletaud and Pilvin, 1994). Especially, the simulation and prediction of the yield surface distorsion remain difficult to achieve with classical phenomenological models, and more or less complicated improvements of these models do not give total satisfaction (Vincent et al., 2002; Kuroda and Tvergaard, 2001).

Polycrystalline models generally give a good description of the plastic anisotropy and yield surface distorsion, because they are based intrinsically on slip lattice as well as texture informations (Cailletaud, 1992; Pilvin 1990; Calloch, 1997). Nevertheless, their main disadvantage is the large computation times needed for parameter identification and finite element calculations. Indeed, these models are built in such a way that the number of internal variables to compute may be a thousand times larger than the number of internal variables using a phenomenological model. Thus one understands the relatively rare use of polycrystalline models in industrial calculations.

In the field of nuclear industry, important work has been done on the viscoplastic behavior of zirconium alloys, using phenomenological models as well as polycrystalline approaches, both combined to experimental multiaxial tests performed on tubes. These multiaxial tests (pure tension, biaxial tension, tension-torsion) are necessary to investigate the material response to the complex loadings that are applied to the nuclear fuel rods in Pressurized Water Reactors. It can be seen that an accurate description of the anisotropy and irradiation effects requires quite a complex set of equations in phenomenological approaches (Robinet, 1995; Schaeffler, 1997; Richard et al, 2003) while the multi-scale approach allows much simpler developments (Lebensohn and Tomé, 1993 ; Geyer, 1999 ; Onimus, 2003). Nevertheless, none of these models is used at present time in really industrial finite element calculations because of their complexity or too large CPU time.

In order to combine the good predictive ability of polycrystalline approaches and the moderate CPU time necessary to conduct safety analyses for the nuclear industry, we propose in the present paper a simplified polycrystalline model (the so-called RL model - Rousselier and Leclercq, 2004) to simulate the anisotropic viscoplastic behavior of metallic materials. A generic method is presented that makes it possible to build a simplified anisotropic material texture, based on the principal features of the pole figures. The method is applied to recrystallised zirconium alloys. The model aims at representing the mechanical behavior of polycrystals with a minimum number (between 5 and 10) of crystalline orientations ("grains"), and considers only 6 slip systems by grain for describing the inelastic strain tensor, whatever the crystallographic lattice of the material under study is. It must be very clear that our goal is not to represent accurately the material texture, but only its mechanical effect on the response of a structure to a mechanical loading. Nevertheless, we show in the following that our choice of the crystallographic orientations is closely related to the real material texture, and moreover that it can be identified thanks to an optimization algorithm by comparison with the results of experimental tests.

In the next section, we recall the major features of the simplified viscoplastic RL-polycrystalline model. In section 3, we show how the model has been applied to the simulation of the anisotropic behavior of a recrystallized zirconium alloy. Particular attention is paid to the description of the parameter identification of the model using several multiaxial tests. An example of finite element calculation performed on a tube is presented in section 4. Concluding remarks are given in the last section of this paper.

## 2. The simplified viscoplastic RL-polycrystalline model

The model is presented in the case of small deformations and small rotations. We want to emphasize that the RL model does not depart from classical polycrystalline models. Consequently, all existing and future theoretical and software developments apply easily to the model, with regard for example to finite deformations, rotation of crystallographic orientations, localization self-consistent models, or slip systems constitutive equations.

Let $\underline{n}_s$ be the normal vector of a slip plane and $\underline{l}_s$ the unit slip direction vector. The orientation tensor of the slip system number $s$ is defined as:

$$\underline{\underline{m}}_s = (\underline{n}_s \otimes \underline{l}_s + \underline{l}_s \otimes \underline{n}_s)/2 \qquad (1)$$

We consider only 6 universal slip systems corresponding to shear and extension strain rates in the local frame $x_1 x_2 x_3$ of a given crystallographic orientation (Figure 1):

$$\underline{n}_1 = \begin{bmatrix} 1 & 0 & 0 \end{bmatrix}, \ \underline{n}_2 = \begin{bmatrix} 0 & 1 & 0 \end{bmatrix}, \ \underline{n}_3 = \begin{bmatrix} 0 & 0 & 1 \end{bmatrix}$$

$$\underline{l}_1 = \begin{bmatrix} 0 & 1 & 0 \end{bmatrix}, \ \underline{l}_2 = \begin{bmatrix} 0 & 0 & 1 \end{bmatrix}, \ \underline{l}_3 = \begin{bmatrix} 1 & 0 & 0 \end{bmatrix}$$

$$\underline{\underline{m}}_1 = \frac{1}{2}\begin{bmatrix} 0 & 1 & 0 \\ 1 & 0 & 0 \\ 0 & 0 & 0 \end{bmatrix}, \ \underline{\underline{m}}_2 = \frac{1}{2}\begin{bmatrix} 0 & 0 & 0 \\ 0 & 0 & 1 \\ 0 & 1 & 0 \end{bmatrix}, \ \underline{\underline{m}}_3 = \frac{1}{2}\begin{bmatrix} 0 & 0 & 1 \\ 0 & 0 & 0 \\ 1 & 0 & 0 \end{bmatrix}$$

$$\underline{n}_4 = \begin{bmatrix} 1/\sqrt{2} & 1/\sqrt{2} & 0 \end{bmatrix}, \ \underline{n}_5 = \begin{bmatrix} 0 & 1/\sqrt{2} & 1/\sqrt{2} \end{bmatrix}, \ \underline{n}_6 = \begin{bmatrix} 1/\sqrt{2} & 0 & 1/\sqrt{2} \end{bmatrix}$$

$$\underline{l}_4 = \begin{bmatrix} -1/\sqrt{2} & 1/\sqrt{2} & 0 \end{bmatrix}, \ \underline{l}_5 = \begin{bmatrix} 0 & -1/\sqrt{2} & 1/\sqrt{2} \end{bmatrix}, \ \underline{l}_6 = \begin{bmatrix} 1/\sqrt{2} & 0 & -1/\sqrt{2} \end{bmatrix}$$

$$\underline{\underline{m}}_4 = \frac{1}{2}\begin{bmatrix} -1 & 0 & 0 \\ 0 & 1 & 0 \\ 0 & 0 & 0 \end{bmatrix}, \ \underline{\underline{m}}_5 = \frac{1}{2}\begin{bmatrix} 0 & 0 & 0 \\ 0 & -1 & 0 \\ 0 & 0 & 1 \end{bmatrix}, \ \underline{\underline{m}}_6 = \frac{1}{2}\begin{bmatrix} 1 & 0 & 0 \\ 0 & 0 & 0 \\ 0 & 0 & -1 \end{bmatrix} \tag{2}$$

The incompressible plastic strain rate of a given orientation or "grain" results from the multiple slip:

$$\underline{\underline{\dot{\varepsilon}}}^p = \sum_{s=1}^{6} \underline{\underline{m}}_s \dot{\gamma}_s = \frac{1}{2}\begin{pmatrix} \dot{\gamma}_6 - \dot{\gamma}_4 & \dot{\gamma}_1 & \dot{\gamma}_3 \\ \dot{\gamma}_1 & \dot{\gamma}_4 - \dot{\gamma}_5 & \dot{\gamma}_2 \\ \dot{\gamma}_3 & \dot{\gamma}_2 & \dot{\gamma}_5 - \dot{\gamma}_6 \end{pmatrix} \tag{3}$$

where $\dot{\gamma}_s$ is the slip rate of the slip system $s$.

For hcp crystals, the deformation results from 5 main slip families (for instance, see Geyer, 1999; Starolesky and Anand, 2003): basal B<a>, prismatic P<a>, pyramidal π1<a>, π1<c+a> and π2<c+a> (Figure 2a). Only pyramidal slips <c+a> contribute to the strain rate component $\dot{\varepsilon}_{33}^{p}$. According to equation (3), the systems 5 and 6 of the model will be associated with these physical slip systems. For the same reason, the systems 2 and 3 of the model will be associated with basal slips that contribute to the strain rate components $\dot{\varepsilon}_{13}^{p}$ and $\dot{\varepsilon}_{23}^{p}$. Finally, the systems 1 and 4 of the model will be associated with prismatic slips that contribute to the strain rate components $\dot{\varepsilon}_{11}^{p}$, $\dot{\varepsilon}_{12}^{p}$, and $\dot{\varepsilon}_{22}^{p}$. So, for hcp materials, we should consider 3 slip families (1,4), (2,3) and (5,6) that will have the properties respectively of prismatic, basal and pyramidal slips (Figure 2b). In order to take into account the fact that systems 1 and 4 have not the same schematic representation in Figure 2b, we choose to split the "prismatic family" into two parts, and we assume that they have not the same properties, especially in terms of critical resolved shear stress. We do not pretend that these non physical slip systems accurately represent the physical ones. Again it must be clear that our goal is only to model the macroscopic mechanical behavior of the material.

Any constitutive equation for the slip systems can be used with the RL-polycrystal model, for example the power law form of Asaro and Needleman (1985) or equations depending on dislocation densities (for instance, see Hoc and Forest, 2001; Hoc et al., 2003). In the following, we will use a phenomenological model with two internal variables, isotropic $r_s$ and kinematic $X_s$, as proposed by Cailletaud (1992) :

$$\dot{\gamma}_s = \dot{v}_s Sign(\tau_s - X_s) \qquad (4)$$

$$\dot{v}_s = Max\left[0, \left(\frac{|\tau_s - X_s| - r_s}{K_s}\right)^{n_s}\right] \qquad (5)$$

$$r_s = \tau_s^c + \sum_{t=1}^{6} Q_{st}\left[1 - \exp(-b_t v_t)\right] + Q\left[1 - \exp(-b v_s)\right] \qquad (6)$$

$$\dot{X}_s = a_s \dot{\gamma}_s - c_s X_s \dot{v}_s \qquad (7)$$

Coefficients $Q_{st}$ are the components of the hardening moduli matrix $\underline{\underline{Q}}$. The diagonal terms represent the self-hardening of each system, and the non-diagonal terms the latent hardening. Generally speaking, $\underline{\underline{Q}}$ can be non-symmetric. For simplicity, in the following, we have chosen a symmetric form for $\underline{\underline{Q}}$. This induces 8 different values to take into account the 4 slip families of the hcp structure. There is no latent hardening for the kinematic variables. $\tau_s^e$ is the initial critical value of the resolved shear stress $\tau_s$ on the slip system $s$, given by:

$$\tau_s = \underline{\underline{\sigma}}_g : \underline{\underline{m}}_s \qquad (8)$$

where $\underline{\underline{\sigma}}_g$ is the stress tensor in the "grain" number $g$ (from 1 to $N$). This stress tensor $\underline{\underline{\sigma}}_g$ is evaluated using localization equations. In the same way as for the constitutive equations of the slip systems, several choices can be made. One can refer to the general presentation of the RL model (Rousselier and Leclercq, 2004) to have an overview of what may be used in the framework of a simplified polycrystalline approach. In the present application, we consider the localization equation proposed by Cailletaud (1987, 1992), in which auxiliary strain variables $\underline{\underline{\beta}}_g$ are memorized for each "grain" :

$$\underline{\underline{\sigma}}_g = \underline{\underline{\Sigma}} + \alpha(\underline{\underline{B}} - \underline{\underline{\beta}}_g) \quad \text{with} \quad \underline{\underline{B}} = \sum_{g=1}^{N} f_g \underline{\underline{\beta}}_g \tag{9}$$

$$\alpha = 2\mu(1-\beta) \quad \text{and} \quad \beta = \frac{2}{15}\frac{4-5\nu}{1-\nu}$$

where $\mu$ and $\nu$ are the shear modulus and the Poisson coefficient, respectively. In equation (9), $\underline{\underline{\Sigma}}$ is the macroscopic stress tensor (over the Representative Volume of Element) and $f_g$ the volume fraction of "grain" $g$.

The evolution of $\underline{\underline{\beta}}_g$ is given by a phenomenological kinematic equation, with 2 adjustable parameters $D$ (Cailletaud, 1987) and $D^*$ (Pilvin, 1990):

$$\underline{\underline{\dot{\beta}}}_g = \underline{\underline{\dot{\varepsilon}}}_g^p - D\left(\underline{\underline{\beta}}_g - D^* \underline{\underline{\varepsilon}}_g^p\right) J_2\left(\underline{\underline{\dot{\varepsilon}}}_g^p\right)\frac{2}{\sqrt{3}} \tag{10}$$

where $J_2\left(\underline{\underline{T}}\right) = \left(\underline{\underline{T}}_d : \underline{\underline{T}}_d / 2\right)^{1/2}$ and $\underline{\underline{T}}_d$ is the deviator of a tensor $\underline{\underline{T}}$ .

In the present study, the parameter $D^*$ is chosen to be zero.

Let us note the phenomenological character of the present scale transition rule. The idea behind the formulation is that the development of a plastic accommodation will decrease the level of the residual intergranular stresses. The shape of the expression is imported from the classical macroscopic models. The original authors (Pilvin and Cailletaud, 1990) suggest to calibrate the parameters of the rule on finite element computations of a representative material element.

This has not been made in the present study where, following the spirit of the model proposed in this paper, we have decided to consider the variables in the set of parameters to be identified. Nevertheless, the values found in the direct identification process are in good agreement with classical values of these parameters.

## 3. Application to the simulation of the recrystallized Zircaloy behavior

### 3.1. Material

The material under study consists in Zircaloy-4 tubes, the weight chemical composition of which is reported in Table 1.

The material is tested in the α-recrystallized state, i.e. after processing, the tubes are submitted to a heat treatment of 700°C ± 30°C during 4 to 5 hours. After recrystallization, the microstructure consists in small equiaxed grains (2 to 5 μm) which are bounded with larger ones (15 to 18 μm) (Bouffioux, 2004).

The reference frame $X_1 X_2 X_3$ of the material is the frame $r\theta z$ of the tubes. The texture of these tubes is strongly anisotropic, see Figure 3 (Robinet, 1995). It shows $\theta z$ pole figure { 0002 }, i.e. basal poles corresponding to axis c of the crystal or axis $x_3$, and pole figure { $10\bar{1}0$ }, i.e. prismatic poles corresponding to axis $x_1$ in RL model.

Note that the texture presented here results in the combination of two effects : a material one (due to the hcp lattice, and to the heat treatment), and an effect due to tube processing. We have to point out that is texture is characteristic of tubes made of recrystallized zirconium alloys (tubes made of cold-word stress relieved material would have significantly different pole figures).

### 3.2. Experimental data base and parameter identification

Several experimental tests have been performed by Robinet (1995) on tube specimens : tension, tension-shear (with several biaxiality ratios $E_{\theta z}/E_{zz}$), and tension-internal pressure (with several biaxiality ratios $E_{\theta\theta}/E_{zz}$). The experiments are performed at room temperature.

In order to proceed to the parameter identification and to the model validation, we decide to split the data base into two parts. The identification data base is built with three tests : two tension tests at two different macroscopic strain rates ($\dot{E} = 6,6.10^{-4}\,s^{-1}$ and $\dot{E} = 6,6.10^{-7}\,s^{-1}$), and one pure shear test (equivalent strain rate $\overline{\dot{E}} = J_2(\underline{\dot{E}}) = 6,6.10^{-4}\,s^{-1}$). The validation data base (i.e. the data base that is used to demonstrate the predictive ability of the model) is built with the remaining experimental tests.

Usually three types of parameters are identified when one uses a polycrystalline approach :

- the material parameters linked with the slip properties of each system (critical resolved shear stress, viscosity, hardening,…) ;

- the matrix defining self- and latent- hardening ($Q_{st}$) ;

- the parameters of the localization equation ($D$ and $\alpha$).

We have considered $\alpha$ to be a parameter, instead of using its theoretical value $\alpha = 2\mu(1-\beta)$, Eq. (9)

.

In the usual case, the texture of the material is an input data which is made of a certain number of grains represented by their crystallographic orientation. Generally, the number of grains is large. For example, Geyer (1999) used 240 orientations to describe a Representative Elementary Volume of Zircaloy-4.

In our case, we assume that it is possible to represent the material texture of the recrystallized zirconium alloy under study with only 6 grains. It is worth to note that this number of 6 grains must be considered as a minimum, in the spirit of the present model to drastically reduce the number of internal variables handled by finite element codes. Using more grains (in agreement with the experimental evidences of the pole figures), would lead to improve the global responses of the model, as was shown by Geyer (1999) with 240 grains. But the price to pay in terms of CPU time is not worth compared to the relatively low gain in terms of mechanical response.

The Euler angles of our 6 grains are reported in Table 2. One can see that three constants are not assigned ($\psi_a$, $\varphi_b$, $\varphi_c$). This is an original feature of our approach to consider these constants as parameters, and to *incorporate them into the parameter identification procedure.*

Of course, the 6 grains described above are relevant for (and only for) recrystallyzed hcp materials that have pole figures similar to those of Figure 3. For other materials (e.g. cold-work stress relieved materials) with different types of texture, the Euler angles presented in Table 2 are no more relevant and should be re-estimated.

Figure 4a presents the pole figures (axis x3-left and axis x1-right) of the six grains defined in Table 2, with $\psi_a$, $\varphi_b$, $\varphi_c$ as parameters. One can thus figure out how the 3 angles act on the position of the poles to define the material texture. Identifying these parameters gives us the opportunity to verify the ability of the model to match not only with macroscopic tests, but also with the real material texture.

Thus, the parameters of the RL-polycrystal model have been identified using the identification data base defined above. The parameter values are reported in Table 3. As mentioned above, four slip families are considered (i.e., 1, 4, 2-3 and 5-6), each family having its own set of parameters. The self- and latent- hardening matrix has the following form :

$$Q = \begin{pmatrix} Q_1 & Q_{12} & Q_{12} & Q_{14} & Q_{15} & Q_{15} \\ Q_{12} & Q_2 & Q_{23} & Q_{12} & Q_{15} & Q_{15} \\ Q_{12} & Q_{23} & Q_2 & Q_{12} & Q_{15} & Q_{15} \\ Q_{14} & Q_{12} & Q_{12} & Q_1 & Q_{15} & Q_{15} \\ Q_{15} & Q_{15} & Q_{15} & Q_{15} & Q_5 & Q_{56} \\ Q_{15} & Q_{15} & Q_{15} & Q_{15} & Q_{56} & Q_5 \end{pmatrix} \tag{11}$$

One has to point out that, even if the RL slip families are not physical ones, the identified values of the critical resolved shear stresses are in agreement with what is classically observed on non-irradiated zirconium alloys, i.e.

$$\tau_c \text{ prismatic} < \tau_c \text{ basal} < \tau_c \text{ pyramidal}$$

The numerical values obtained for the whole set of parameters are consistent with the ones obtained with standard polycrystalline approaches on the same type of material (see for instance Geyer, 1999).

The comparisons between experiments and simulations of the identification data base are presented on Figs. 5 and 6. The simple lines present RL model simulations. The bold lines correspond to simulations performed with the use of a classical macroscopic viscoplastic model (using one kinematic and one isotropic hardening, see for instance Lemaitre and Chaboche, 1985) in which the yield surface is defined by a Hill criterion, in order to take into account the Zircaloy anisotropy. Of course, this model has been identified with the same data base as the RL model. Use of this model is made in order to have a comparison basis between the approach proposed in this paper and more classical models still widely used. The simulated results are good, and cannot allow one to distinguish between the two approaches.

On Figure 4b, one can see the pole figures (axis x3-left and axis x1-right) of the 6 grains representing the texture of the material under study, after identification of the 3 constants $\psi_a$, $\varphi_b$, $\varphi_c$. These pole figures are to be compared with the ones presented on Figure 3. The good agreement observed here is a first validation of our choice to integrate 3 Euler angles into the parameter identification procedure (see Table 2). A second validation of this original choice will follow with the study of the predictive ability of the RL model applied to the recrystallized Zircaloy-4.

### 3.3 Predictive ability of the model and discussion

In order to investigate its predictive ability, the model has been tested on several loading situations which had not been used in the identification data base. These multiaxial tests are conducted at the same equivalent (in the von Mises sense) strain rate ($\bar{\bar{E}} = 6,6.10^{-4}\,s^{-1}$). We present here three tension-shear tests with three different strain biaxiality ratios ($E_{\theta z}/E_{zz}$), and two tension-internal pressure tests, with two different strain biaxiality ratios ($E_{\theta\theta}/E_{zz}$). The simulated results are compared with the experimental ones on Figures 7 to 11.

Several features are to be pointed out:

i.      First of all, we have to precise that our goal was to test the RL model as concerns its ability to represent the *anisotropic* features of the material in terms of yield stress and flow direction. To this aim, we have used only constant strain rate tests to identify the model parameters. As a matter of fact we do not, in the following, draw any conclusion on the predictive ability of the response of the model to other mechanical loadings cyclic tests, creep or relaxation tests, … Particularly, we are aware that only two experimental tests are not sufficient enough to identify the model parameters associated with all the material characteristics.

ii.      As concerns the RL model, agreements between simulation and experiments are good, for several loading situations and especially the tension-internal pressure tests which were not represented in the identification data base. This confirms the ability of the model to represent situations which have not been taken into account in the identification procedure, especially as concerns anisotropy. We would like to point out that getting this result was not obvious, because of the small number of grains used to simulate the texture of the Zircaloy-4 tube.

iii.   An important feature, linked with the previous one, is that the RL model can be totally identified with only two experimental tests, representing two loading directions (pure tension $zz$, and pure shear $\theta z$). A confirmation of this can be found on Figures 12 and 13, where the critical stresses of the 6 slip systems are represented for each grain (from grain 1 to grain 6), in the macroscopic $zz\text{-}\theta\theta$ and $zz\text{-}z\theta$ planes. It can be seen here that the use of a tension test conducted up to minimum 500 MPa allows to activate and thus identify the material parameters of slip systems # 1, 4, 5 and 6. Note that in the case of a tensile test, slip systems # 2 and 3 are never activated in the RL model. It is interesting to remember that these two systems are associated to basal slip, which is known to be rarely observed when studying the experimental mechanisms of unirradiated Zircaloy-4 (Geyer, 1999). Using a pure shear test conducted up to 250 MPa allows one to activate slip systems # 1, 4 and 3. As the material parameters of slip system # 2 are equal to those of slip system # 3, it is therefore demonstrated that every slip system has been activated at least once in the simulation of the identification data base. This explains why these two testing directions are sufficient for the parameter identification, provided the experiments are conducted far enough in terms of strain (and thus stress) value. The activation of many slip systems at $\Sigma_{\theta\theta} \approx 500$ MPa in Figure 12 explains why the pressure curves are so flat in Figures 10 and 11, a feature that is not due to curve fitting but to the texture of the material and cannot be predicted by macroscopic models.

iv.    The comparison with the macroscopic approach shows the classical feature that the Hill yield surface cannot be identified without appropriate experimental tests in the three directions. Typically, our identification data base does not include any internal pressure test, and this results in the impossibility of identifying any anisotropy between the $rr$, $\theta\theta$, and $zz$ directions for the macroscopic model, as shown on Figs. 10 and 11. As a matter of fact, the identification data base makes it possible to identify only the shear ($\theta z$) anisotropy for this kind of model.

v.    The present study shows that, in the case of an anisotropic material such as Zircaloy, the parameters of the RL model can be identified knowing the general features of the material texture, and with experimental tests performed in two directions (pure tension and pure shear) provided these tests are conducted sufficiently far so as to activate each slip system. With these informations, it is possible to get material parameters that are robust enough to ensure a good predictive ability, which is not the case for macroscopic approaches. Of course, enlarging the identification data base with tests performed in other directions allows to improve the parameter identification of both kind of models, but at a higher cost. Once again, introduction of other experimental tests would be essential to identify parameters linked with specific material features such as creep or cyclic effects, which are not the concern of this paper.

Another way to verify the robustness of the material parameters (and of the proposed constitutive equations) is to investigate whether they are dependent on the identification data base or not. In order to answer this question, we have added to the identification data base the equibiaxial tension with internal pressure test ($E_{\theta\theta}/E_{zz}=1$). With the use of this extended data base, we have identified the material parameters of the RL model from the same initial values as with the first identification. The results are presented in able 4. It is remarkable to point out how the values of these new parameters are close to those presented in Table 3. This feature is another proof that a relatively poor identification data base is sufficient to get robust material parameters of the RL model, and that this model is intrinsically predictive.

## 4. A finite element calculation

As the principal purpose of this paper is to propose a material model based on the classical polycrystalline approach and compatible with finite element calculations (i.e., consuming "reasonable" CPU time), we present in the following the simulation of a Zircaloy-4 tube submitted to an internal pressure and an imposed axial displacement. The structure is calculated using both RL model and the Chaboche+Hill macroscopic approach. We compare the numerical results in terms of strain-stress curves, as well as the CPU time needed to perform the calculation.

The 3D mesh of the finite element calculation is presented Figure 14. It consists in a ring, part of a Zircaloy-4 tube, meshed with 200 quadratic elements with 20 nodes. The imposed boundary conditions are such that the axial strain of the ring is zero, and the hoop strain rate is equal to 6,6 $10^{-4}$ s$^{-1}$. Two finite element calculations have been performed, with the Chaboche+Hill approach and the RL model. The macroscopic strain and stress states in the inner part of the tube are reported Figure 15, compared with the experimental data. The results are consistent with the ones shown on Figure 11, which shows the 0D (representative elementary volume) simulation of this test. It can be pointed out that, especially for the tensile results in Figures 11 and 15, the smaller strain behavior is better predicted with the Chaboche+Hill approach, and the RL model fits better for the larger strains. One explanation for this may be found in the fact that the Chaboche+Hill approach is written in a purely viscoplastic way (the viscoplastic strain always exists, even if it is very small for small total strains). On the contrary, the RL model is written in such a way that viscoplastic slip is activated on a given system if and only if the critical value of the resolved shear stress is reached. Thus non elastic strain is intrinsically smaller in the RL approach than in the classical one. At higher strains, the problem is different : in the Chaboche+Hill model the 7 (one tensorial kinematic and one scalar) hardening internal variables saturate rapidly (relatively low strain hardening) and in the RL model the possible activation of 36 (6 grains x 6 slip systems) slip systems makes it possible to have a better description of the global material behavior.

For complete understanding, we have to note that the hoop stress gradient between the inner and the out part of the tube (tube thickness = 570 micrometers, inner radius = 480 millimeters) is approximately 80 MPa. This confirms that the tube may be considered as a *real structure*, and not as a *thin tube*.

Table V presents the CPU times of the two finite element calculations. One can see that CPU time of RL model is not much larger than the one of Chaboche+Hill model (ratio of approximately 5), compared with the accuracy of the simulated results. If one recalls that the Chaboche+Hill model used in the present study needs to integrate 13 internal variables (the elasticity tensor, one tensorial kinematic hardening, one scalar isotropic hardening, the equivalent viscoplastic strain), and that the RL model requires the integration of 48 internal variables, the ratio of 5 in the CPU times obtained in Table 5 can be easily related to the ratio of the number of internal variables to integrate. It must also be pointed out that the same finite element calculation performed with a standard polycrystalline model (for instance Geyer, 1999, with 7206 internal variables to integrate) would have last approximately 200 times longer than the one with RL model.

## 5. Concluding remarks

In the present paper, a simplified polycrystalline model (the so-called RL model) is proposed to simulate the anisotropic viscoplastic behavior of metallic materials. A generic method is presented that makes it possible to build a simplified anisotropic material texture, based on the principal features of the pole figures. The method is applied to a recrystallised zirconium alloy, used as clad material in the fuel rods of nuclear power plants. The objective of this study was to propose a robust, predictive but little CPU time consuming model for industrial finite element calculations.

In order to achieve this goal, several simplifications of the classical polycrystalline approach have been performed : the definition of slip systems and the choice and number of crystalline orientations are the major aspects developed here. As a matter of fact, these simplifications lower the physical basis of the approach proposed in this paper, but it has been shown that the main characteristics of the classical polycrystalline approach were kept : predictive ability and intrinsic way to describe the material anisotropy. Moreover, we have shown that the simplicity of the RL model allows one to follow a parameter identification scheme that can be dedicated to <u>each</u> slip system. The relatively small CPU time (as compared to standard polycrystalline approaches), combined with the possibility to follow each slip system, makes it possible to get robust material parameters with a minimum number of tests in the identification data base.

The finite element calculations performed with the RL model confirm the promising features of this model in terms of good quality of the simulations, and CPU time of the same order as with more classical macroscopic models.

The next step, under progress, is to apply the material parameters of the model to another texture, obtained from recrystallized zirconium alloy sheets, of the same chemical composition. The results, combined with the modeling of zirconium alloys' damage properties, will be exposed in a forthcoming paper.

**Acknowledgements**

The authors would like to thank Dr. Jacques Besson (Centre des Matériaux, Ecole des Mines de Paris) for his fruitful help in the finite element calculations.

**Figure captions**

Figure 1. Representation of the 6 slip systems of the RL model in the local frame.

Figure 2. 2a : the physical slip families of a hcp crystal. 2b : the 4 slip families 1, 4, (2, 3), (5, 6) of a hcp crystal represented in the RL model frame.

Figure 3. The experimental texture of the recrystallized Zircaloy-4. 3a : the $\theta z$ pole figure { 0002 }, i.e. basal poles corresponding to axis c of the crystal. 3b : the $\theta z$ pole figure { $10\bar{1}0$ }, i.e. prismatic poles.

Figure 4. The pole figures of the simulated material (axis x3-left and axis x1-right) 4a : principles of the definition of $\psi_a$, $\varphi_b$, $\varphi_c$. 4b : the pole figures after identification of $\psi_a$, $\varphi_b$, $\varphi_c$.

Figure 5. Identification data base : tension tests. Experimental points, RL model (solid line), Chaboche+Hill (bold line).

Figure 6. Identification data base : pure shear test. Experimental points, RL model (solid line), Chaboche+Hill (bold line).

Figure 7. Simulation data base : tension-shear ($E_{\theta z}/E_{zz} = \sqrt{3}/4$). Experimental points, RL model (solid line), Chaboche+Hill (bold line).

Figure 8. Simulation data base : tension-shear ($E_{\theta z}/E_{zz} = \sqrt{3}/2$). Experimental points, RL model (solid line), Chaboche+Hill (bold line).

Figure 9. Simulation data base : tension-shear ($E_{\theta z}/E_{zz} = \sqrt{3}$). Experimental points, RL model (solid line), Chaboche+Hill (bold line).

Figure 10. Simulation data base : tension-internal pressure ($E_{\theta\theta}/E_{zz} = 1$). Experimental points, RL model (solid line), Chaboche+Hill (bold line).

Figure 11. Simulation data base : tension-internal pressure ($E_{\theta\theta}/E_{zz} = \infty$). Experimental points, RL model (solid line), Chaboche+Hill (bold line).

Figure 12. Critical stresses of the 6 slip systems represented for the material texture, in the macroscopic $zz$-$\theta\theta$ plane.

Figure 13. Critical stresses of the 6 slip systems represented for the material texture, in the macroscopic $zz$-$\theta z$ plane.

Figure 14. 3D mesh of a ring submitted to tension and internal pressure ($E_{\theta\theta}/E_{zz} = \infty$).

Figure 15. Macroscopic strain and stress states at the outer part of the ring. Experimental points, RL model (solid line), Chaboche+Hill (bold line).

Table 1. Weight chemical composition (%) of the Zircaloy-4 under study.

| Cr | Fe | Sn | O | Zr | |
|---|---|---|---|---|---|
| 0,10 | 0,21 | 1,3 | 0,135 | balance | |
| Al (ppm) | C (ppm) | H (ppm) | Hf (ppm) | N (ppm) | Si (ppm) |
| 98 | 151 | 9 | 56 | 24 | 99 |

Table 2. Euler angles of the 6 grains representing the recrystallized Zircaloy-4 texture.

| $g$ | 1 | 2 | 3 | 4 | 5 | 6 |
|-----|---|---|---|---|---|---|
| $\psi_g$ | $\psi_a$ | $\pi - \psi_a$ | $\pi/2$ | $\pi/2$ | $0$ | $0$ |
| $\phi_g$ | $\pi/2$ | $\pi/2$ | $\pi/2$ | $\pi/2$ | $\pi/2$ | $\pi/2$ |
| $\varphi_g$ | $0$ | $0$ | $\varphi_b$ | $-\varphi_b$ | $\varphi_c$ | $-\varphi_c$ |

Table 3. Material parameters of the RL model identified on the recrystallized Zircaloy-4, with the initial identification data base.

| $\psi_a$ | $\varphi_b$ | $\varphi_c$ | | $\alpha$ (MPa) | $D$ |
|---|---|---|---|---|---|
| 54° | 65° | 73,7° | | 37963 | 256 |
| $E$ (MPa) | $\nu$ | | $Q_1$ (MPa) | $Q_2$ (MPa) | $Q_5$ (MPa) |
| 125000 | 0.36 | | 77,4 | 88,8 | 78 |
| $Q_{12}$ (MPa) | $Q_{14}$ (MPa) | $Q_{15}$ (MPa) | $Q_{23}$ (MPa) | $Q_{56}$ (MPa) | |
| 198,7 | 116,3 | 111 | 47,7 | 175 | |
| $Q$ (MPa) | $b$ | | | | |
| 9,8 | 1,14 | | | | |
| $\tau_1^c$ (MPa) | $n_1$ | $K_1$ (MPa) | $\tau_2^c$ (MPa) | $n_2$ | $K_2$ (MPa) |
| 31,4 | 1,34 | 8730 | 227,6 | 1,63 | 5036 |
| $\tau_4^c$ (MPa) | $n_4$ | $K_4$ (MPa) | $\tau_5^c$ (MPa) | $n_5$ | $K_5$ (MPa) |
| 146,2 | 1,12 | 1396,3 | 231,7 | 1,07 | 3463,5 |
| $a_1$ (MPa) | $b_1$ | $c_1$ | $a_2$ (MPa) | $b_2$ | $c_2$ |
| 5400,16 | 5,81 | 920,89 | 2358,34 | 7,77 | 641,28 |
| $a_4$ (MPa) | $b_4$ | $c_4$ | $a_5$ (MPa) | $b_5$ | $c_5$ |
| 5321,46 | 6,33 | 765,4 | 6522,16 | 2,74 | 972,26 |

Table 4. Material parameters of the RL model identified on the recrystallized Zircaloy-4, with the extended identification data base.

| $\psi_a$ | $\varphi_b$ | $\varphi_c$ | | $\alpha$ (MPa) | $D$ |
|---|---|---|---|---|---|
| 57° | 66,7° | 67,3° | | 37633 | 260 |
| $E$ (MPa) | $\nu$ | | $Q_1$(MPa) | $Q_2$(MPa) | $Q_5$(MPa) |
| 125000 | 0,36 | | 78,4 | 89,5 | 76,4 |
| $Q_{12}$(MPa) | $Q_{14}$(MPa) | $Q_{15}$(MPa) | $Q_{23}$(MPa) | $Q_{56}$(MPa) | |
| 195,8 | 116,3 | 112 | 49 | 171,3 | |
| $Q$(MPa) | $b$ | | | | |
| 9,4 | 1,2 | | | | |
| $\tau_1^c$(MPa) | $n_1$ | $K_1$(MPa) | $\tau_2^c$(MPa) | $n_2$ | $K_2$(MPa) |
| 32,4 | 1,37 | 8879 | 231,7 | 1,65 | 5033 |
| $\tau_4^c$(MPa) | $n_4$ | $K_4$(MPa) | $\tau_5^c$(MPa) | $n_5$ | $K_5$(MPa) |
| 141,2 | 1,12 | 1394,1 | 242,1 | 1,07 | 3471,5 |
| $a_1$ (MPa) | $b_1$ | $c_1$ | $a_2$ (MPa) | $b_2$ | $c_2$ |
| 5415,1 | 5,7 | 908,88 | 2343,29 | 7,8 | 628,53 |
| $a_4$ (MPa) | $b_4$ | $c_4$ | $a_5$ (MPa) | $b_5$ | $c_5$ |
| 5262,83 | 6,24 | 765,9 | 6641,55 | 2,69 | 972,26 |

Table 5. CPU times (PC/Linux) obtained with Chaboche+Hill and RL models.

| Model | Hill | RL |
|-------|------|-----|
| CPU time (s) | 196 | 977 |

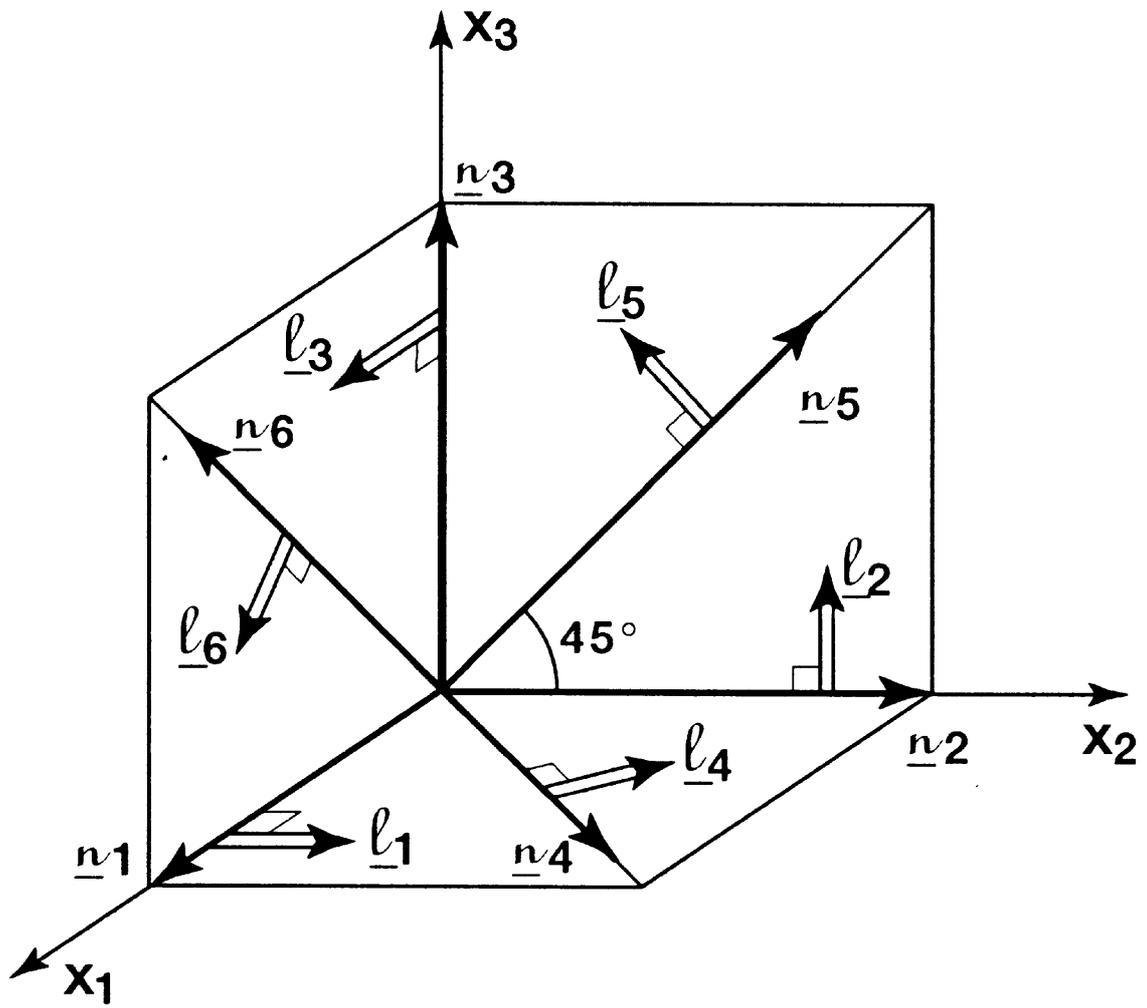

Figure 1

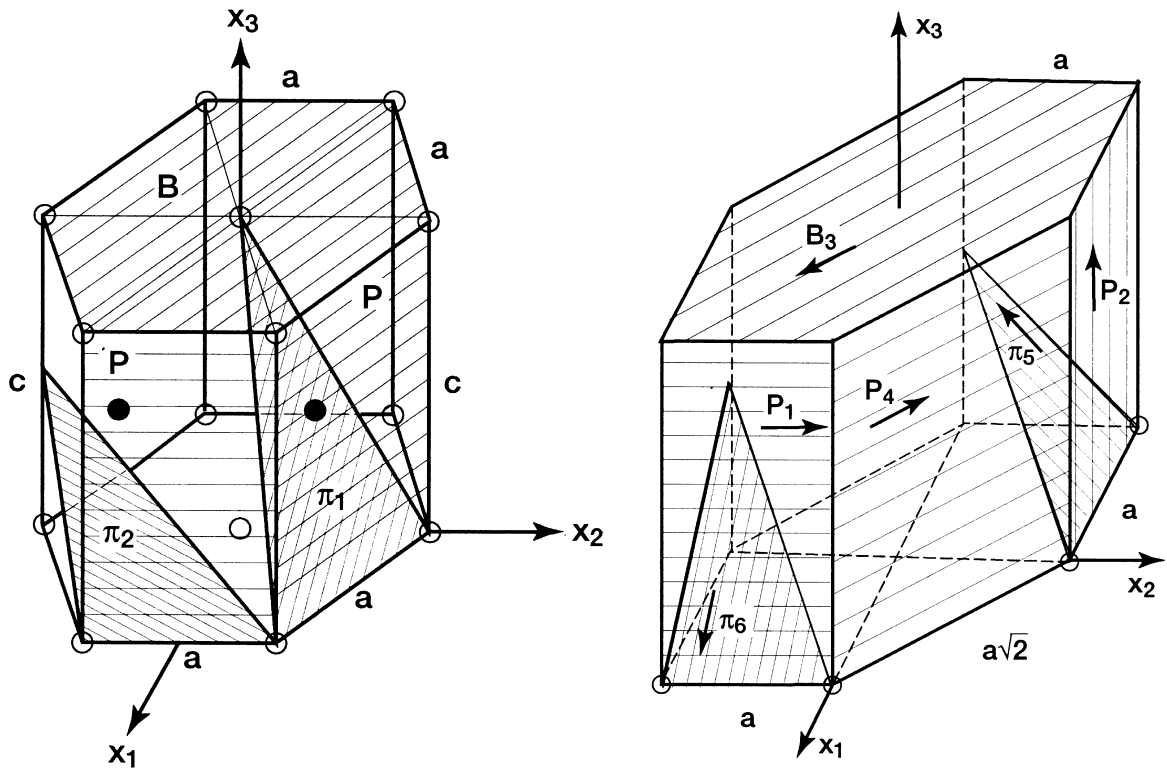

Figure 2

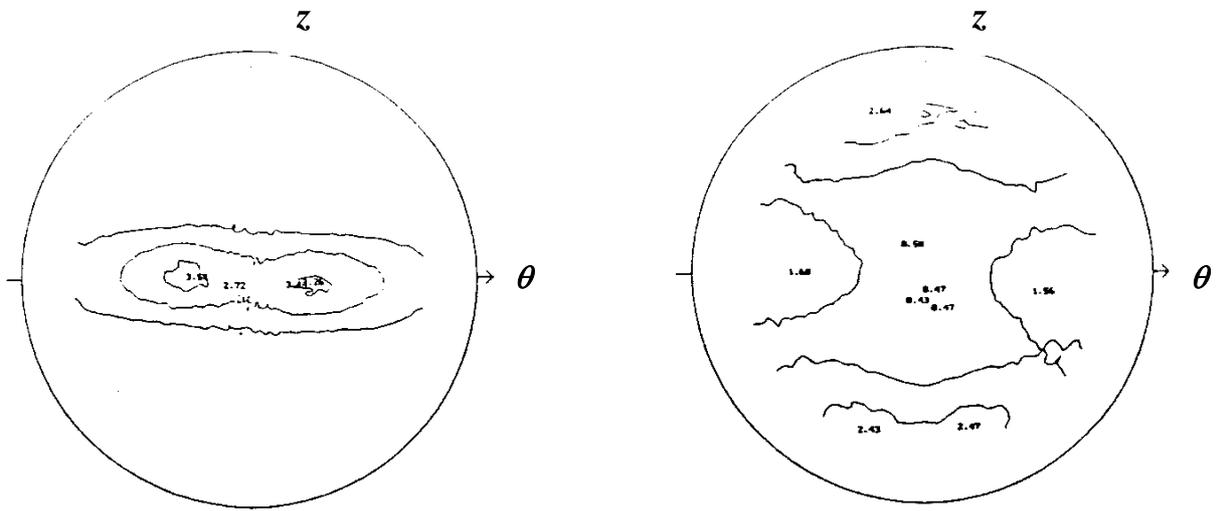

Figure 3

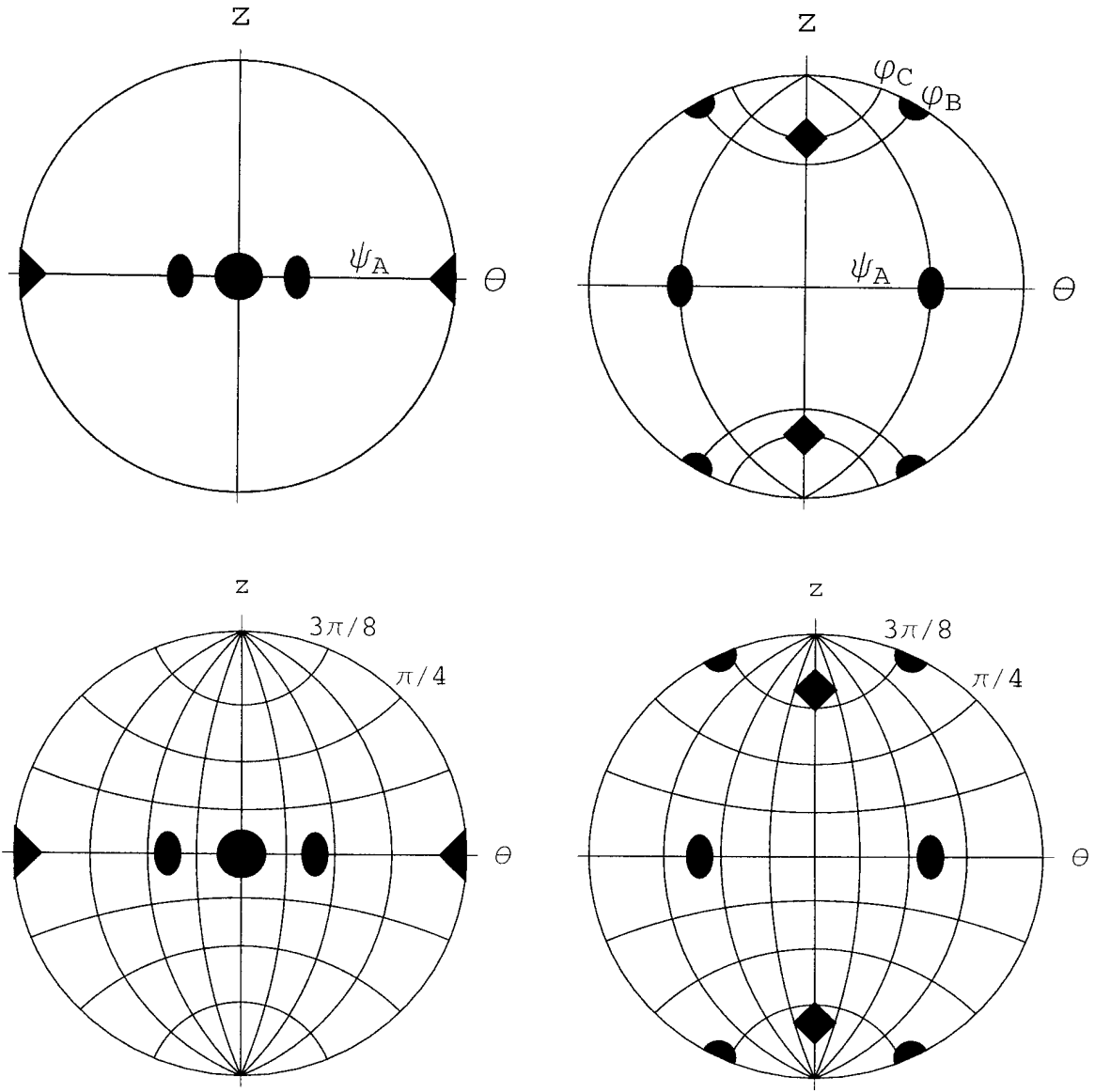

Figure 4a

Figure 4b

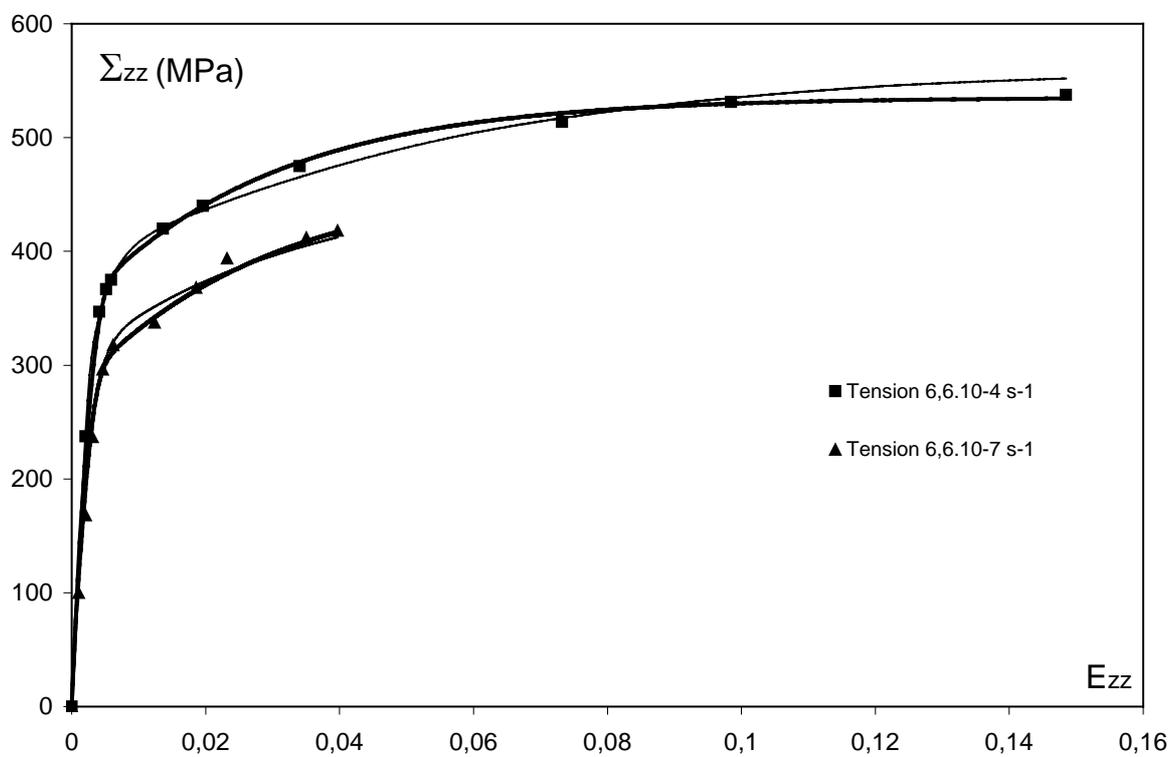

Figure 5

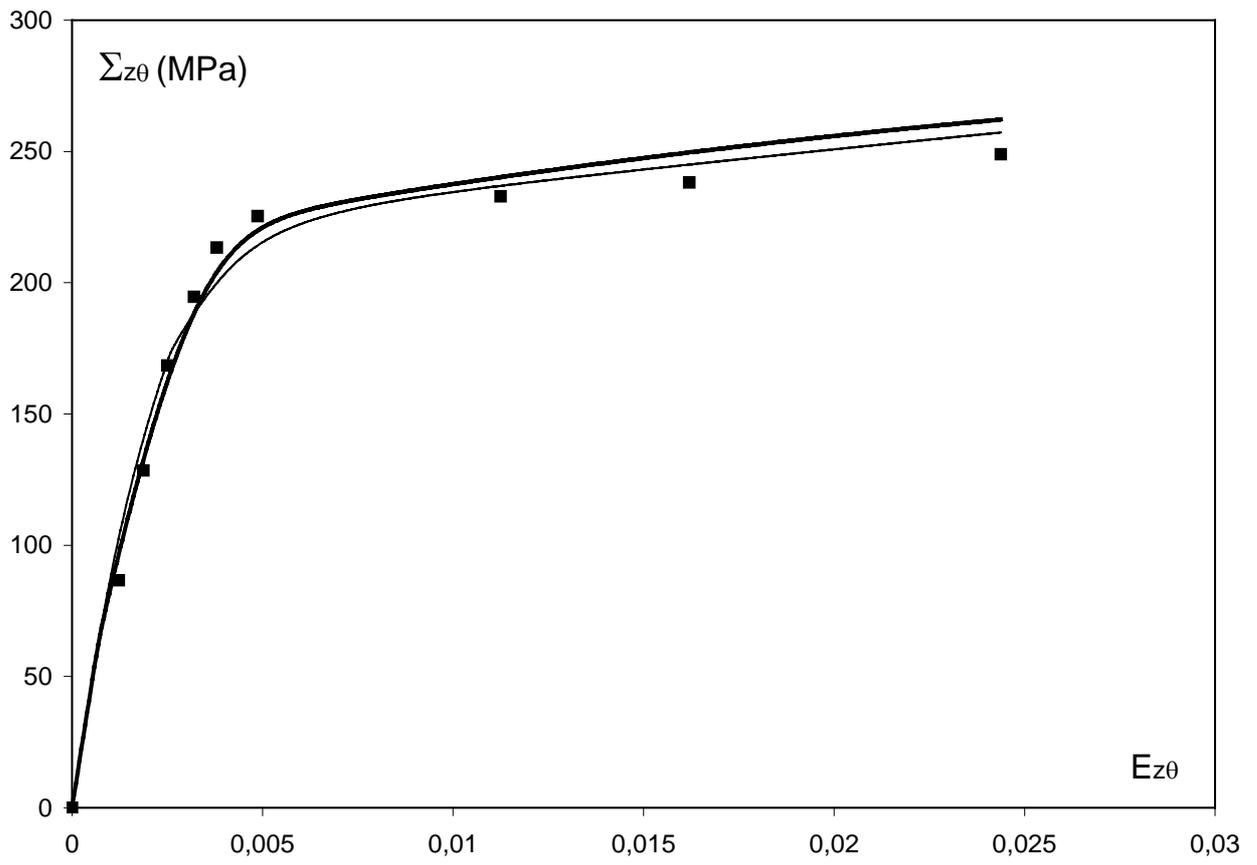

Figure 6

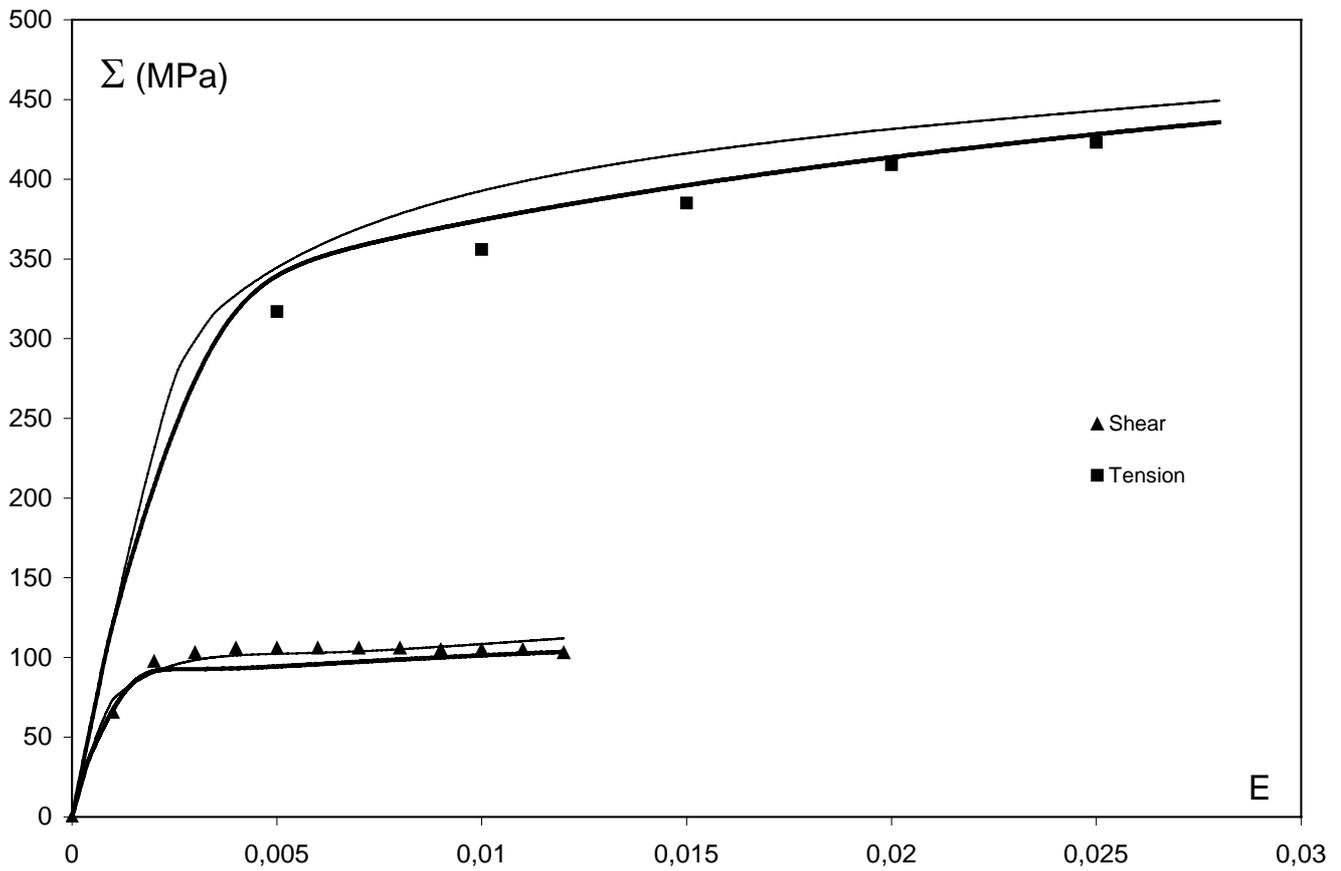

Figure 7

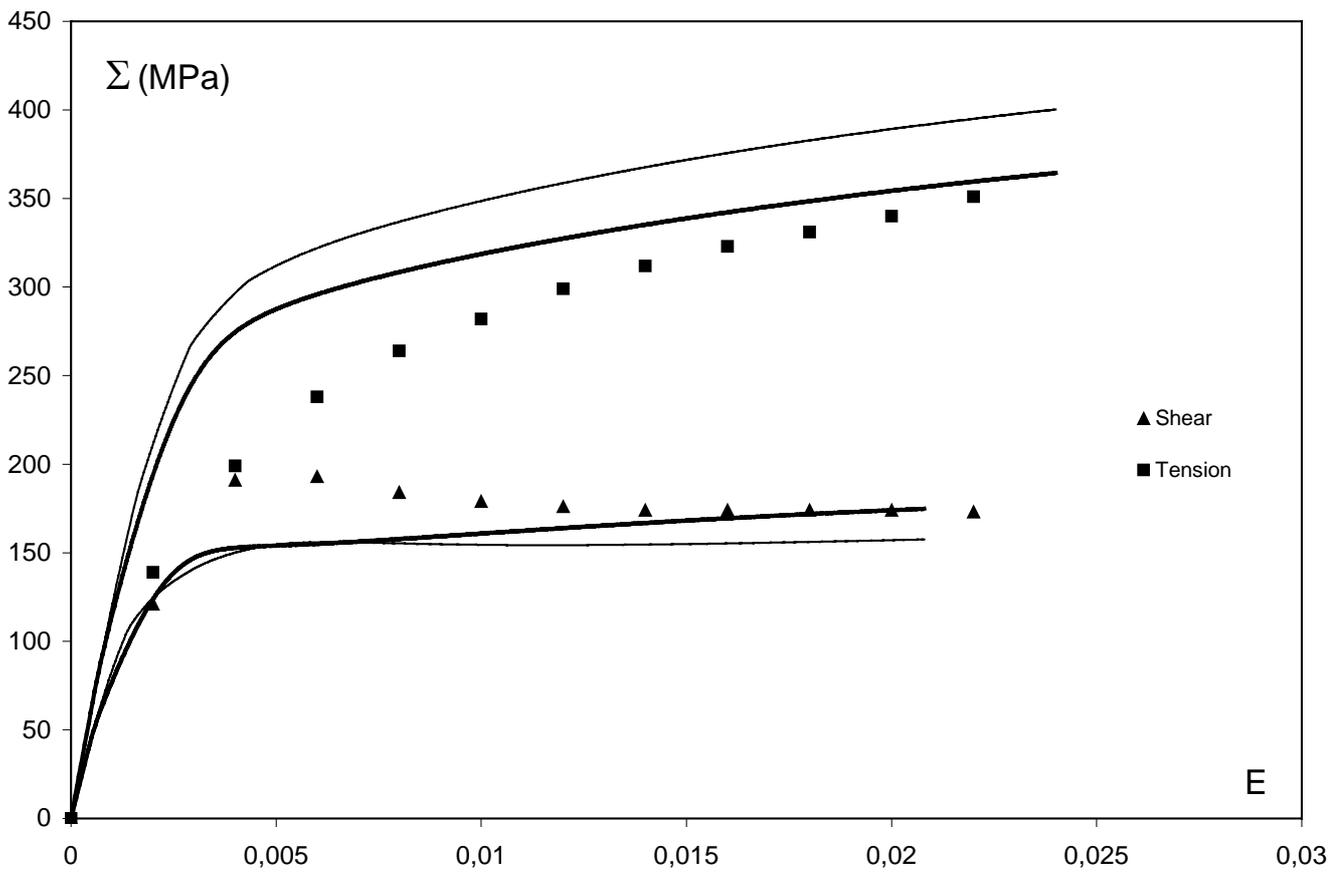

Figure 8

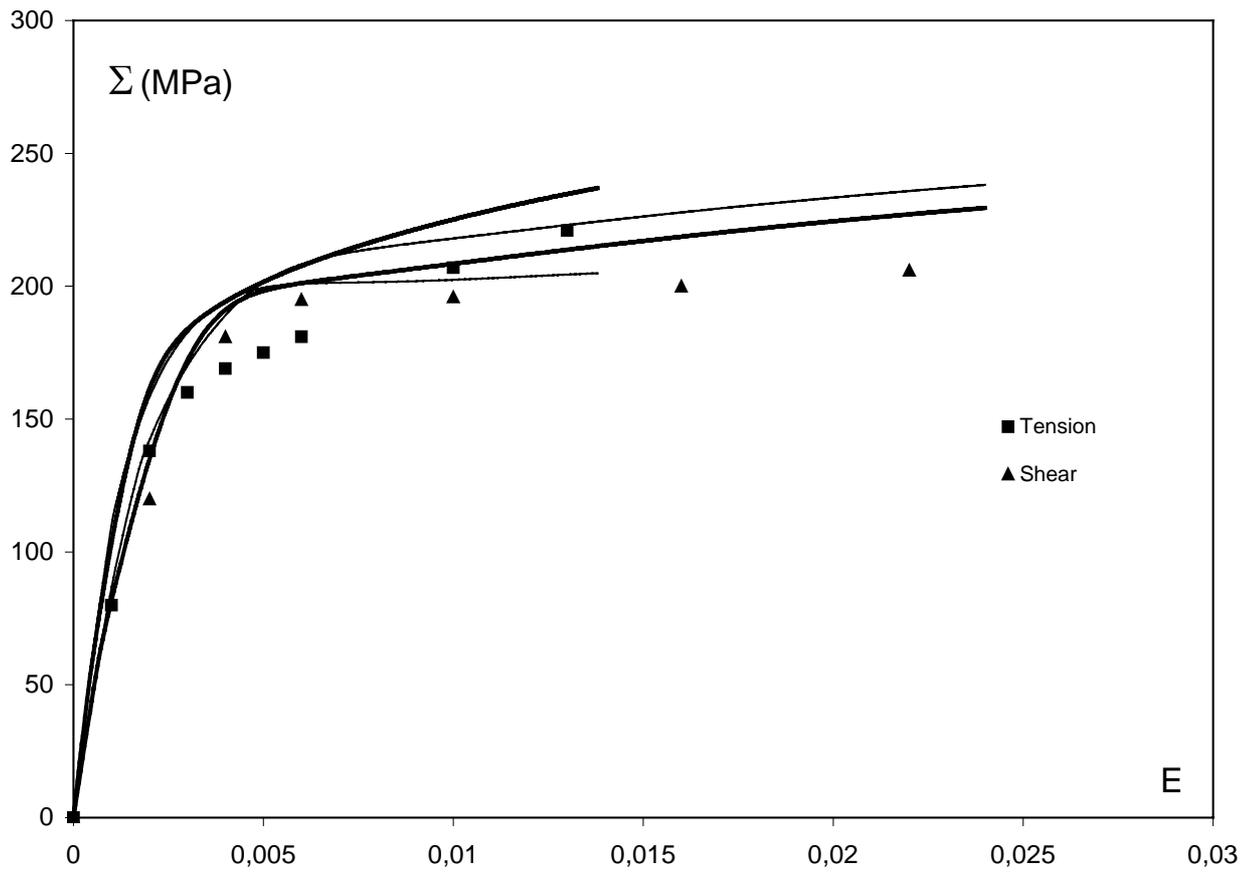

Figure 9

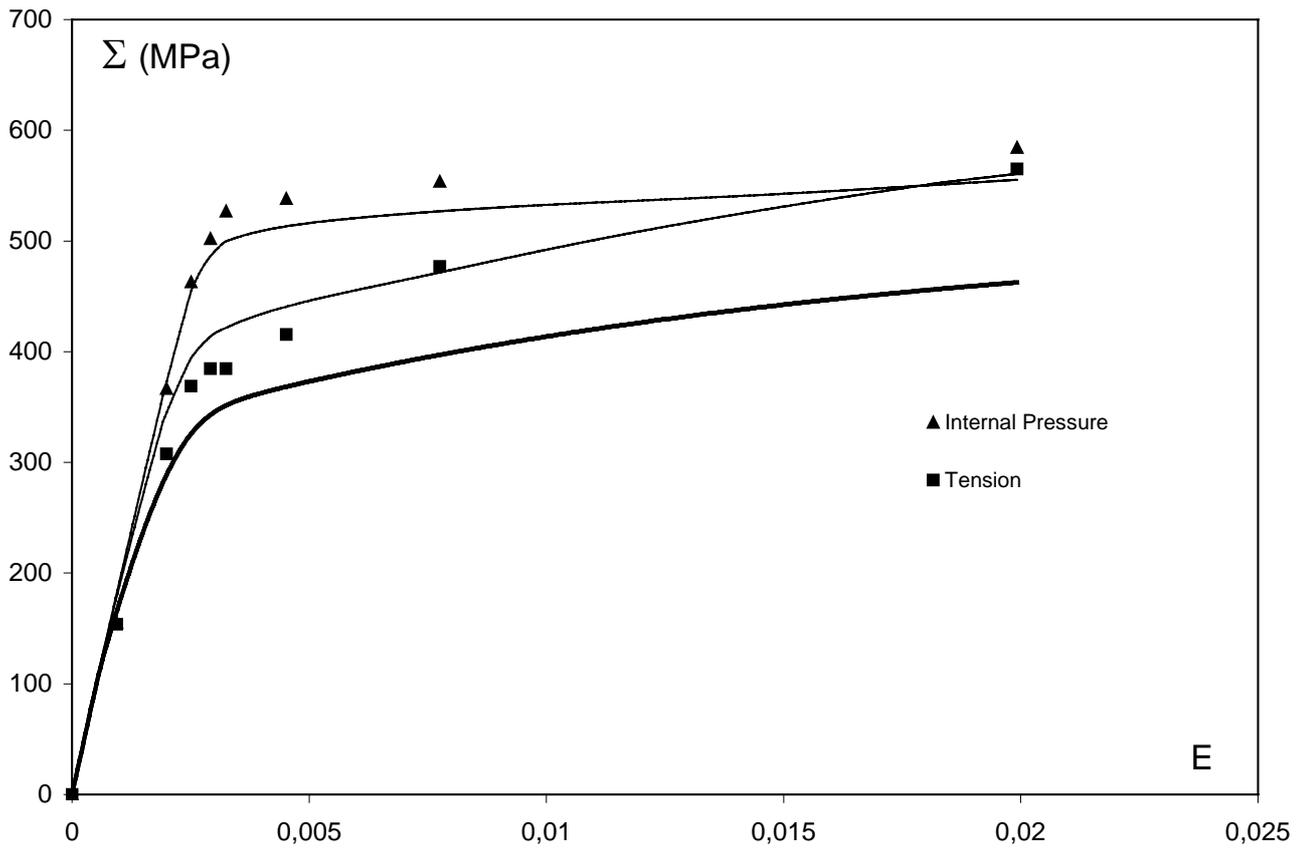

Figure 10

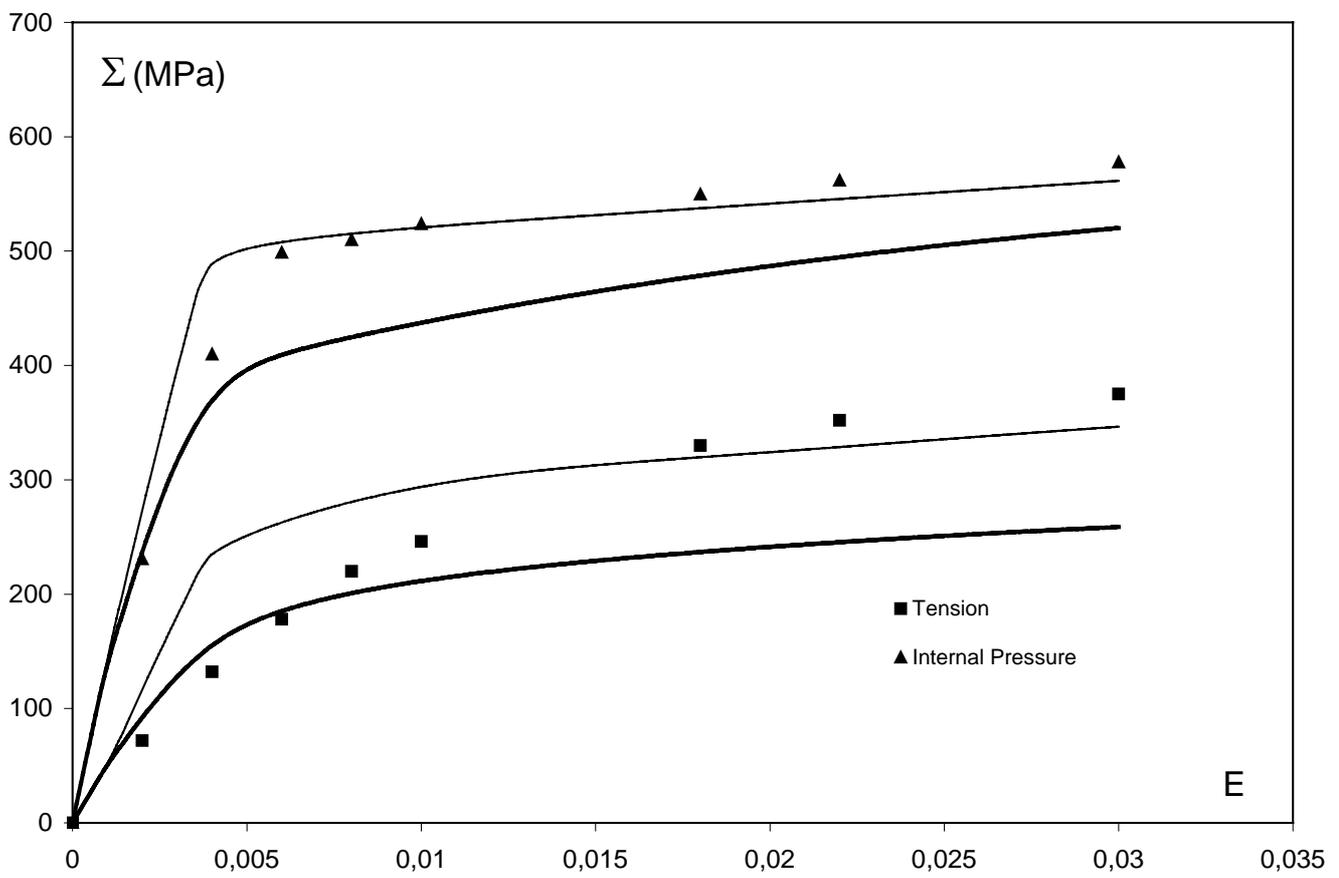

Figure 11

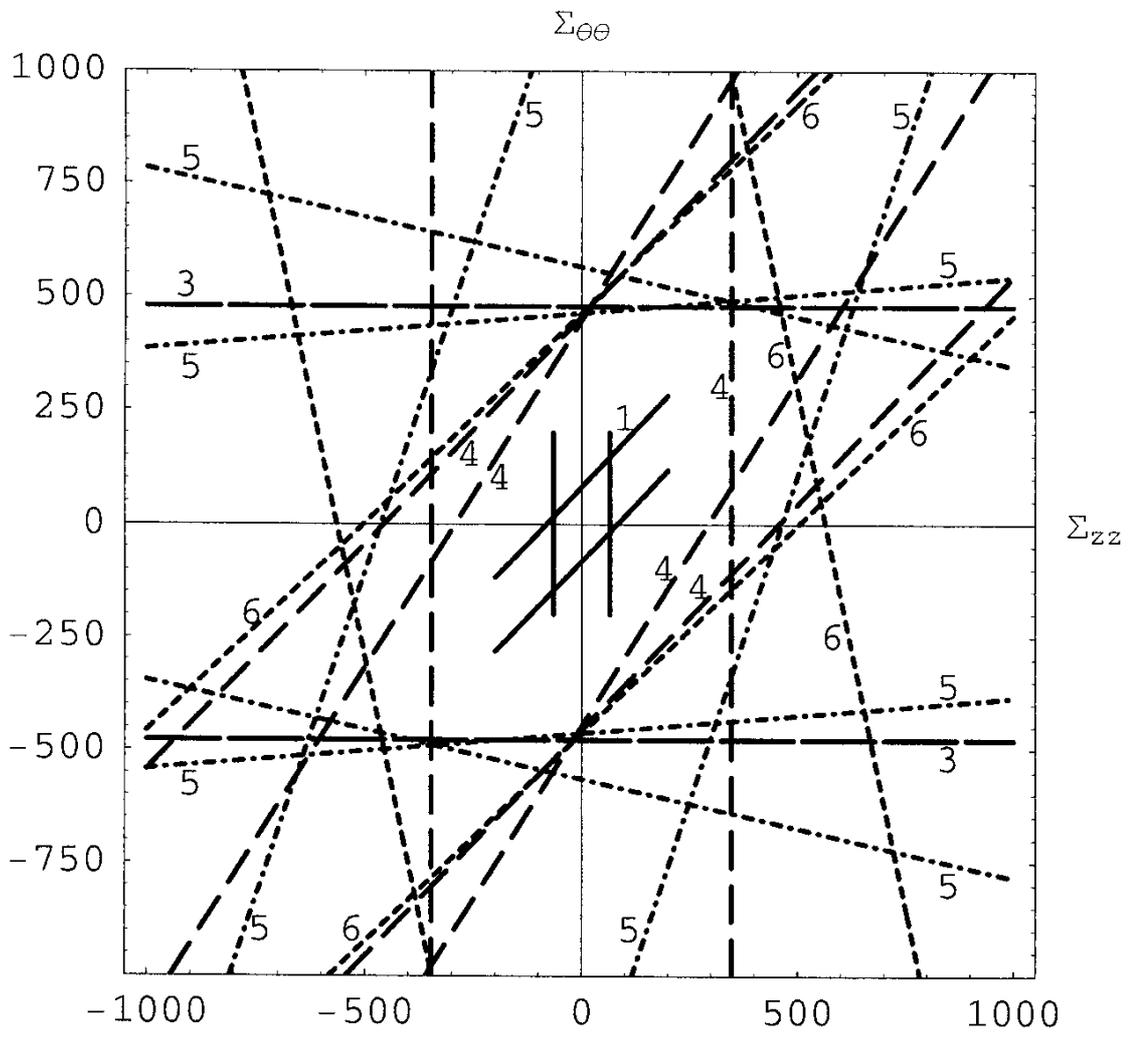

Figure 12

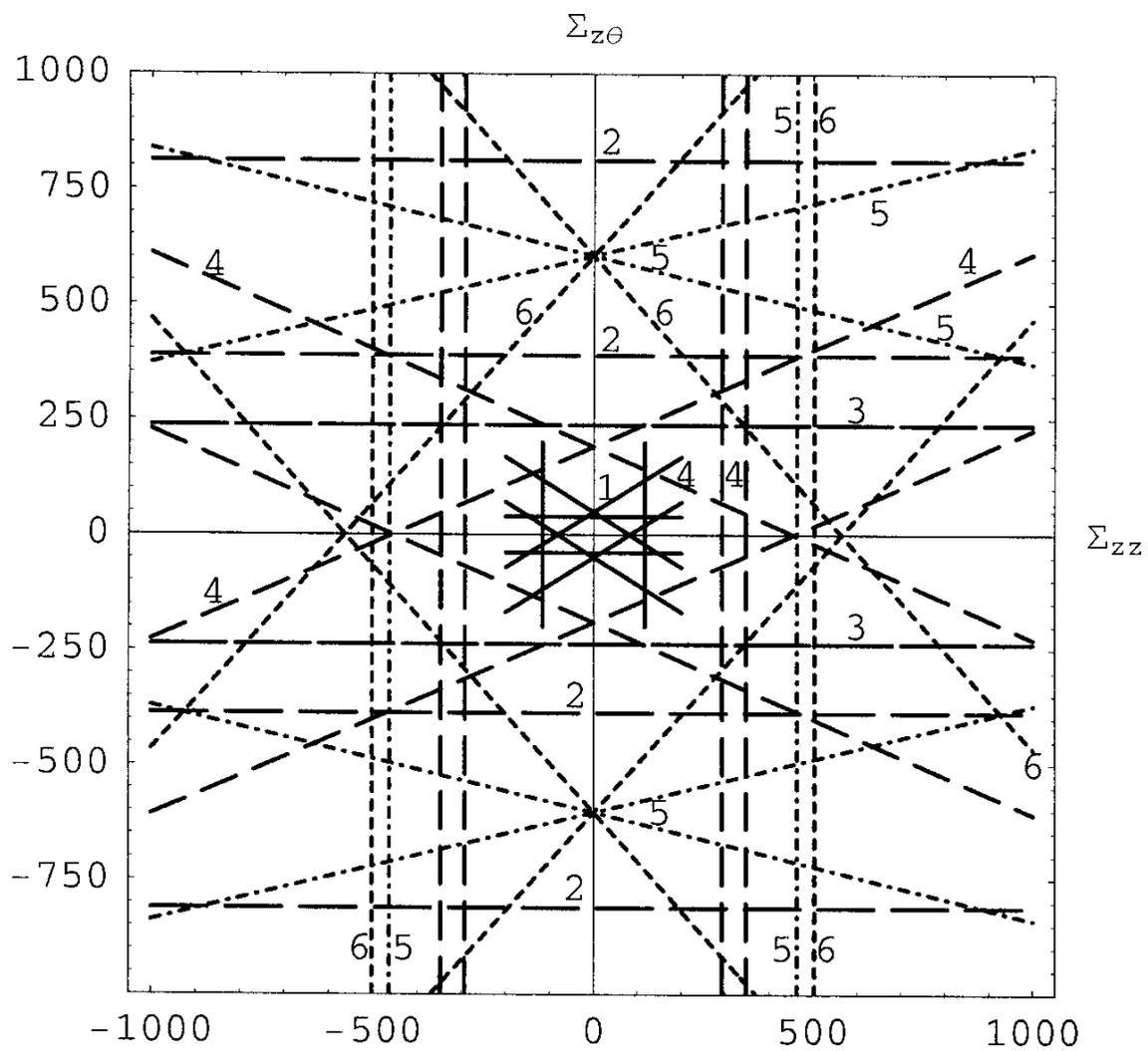

Figure 13

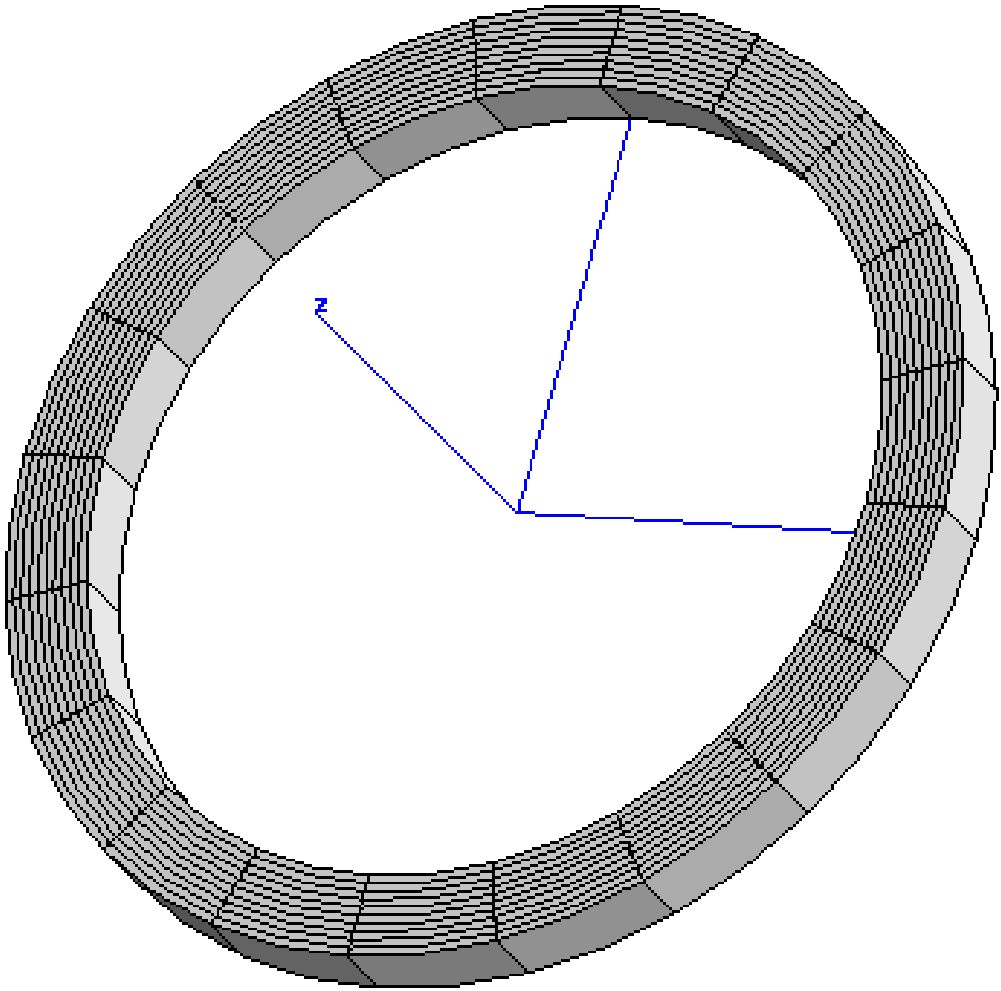

Figure 14

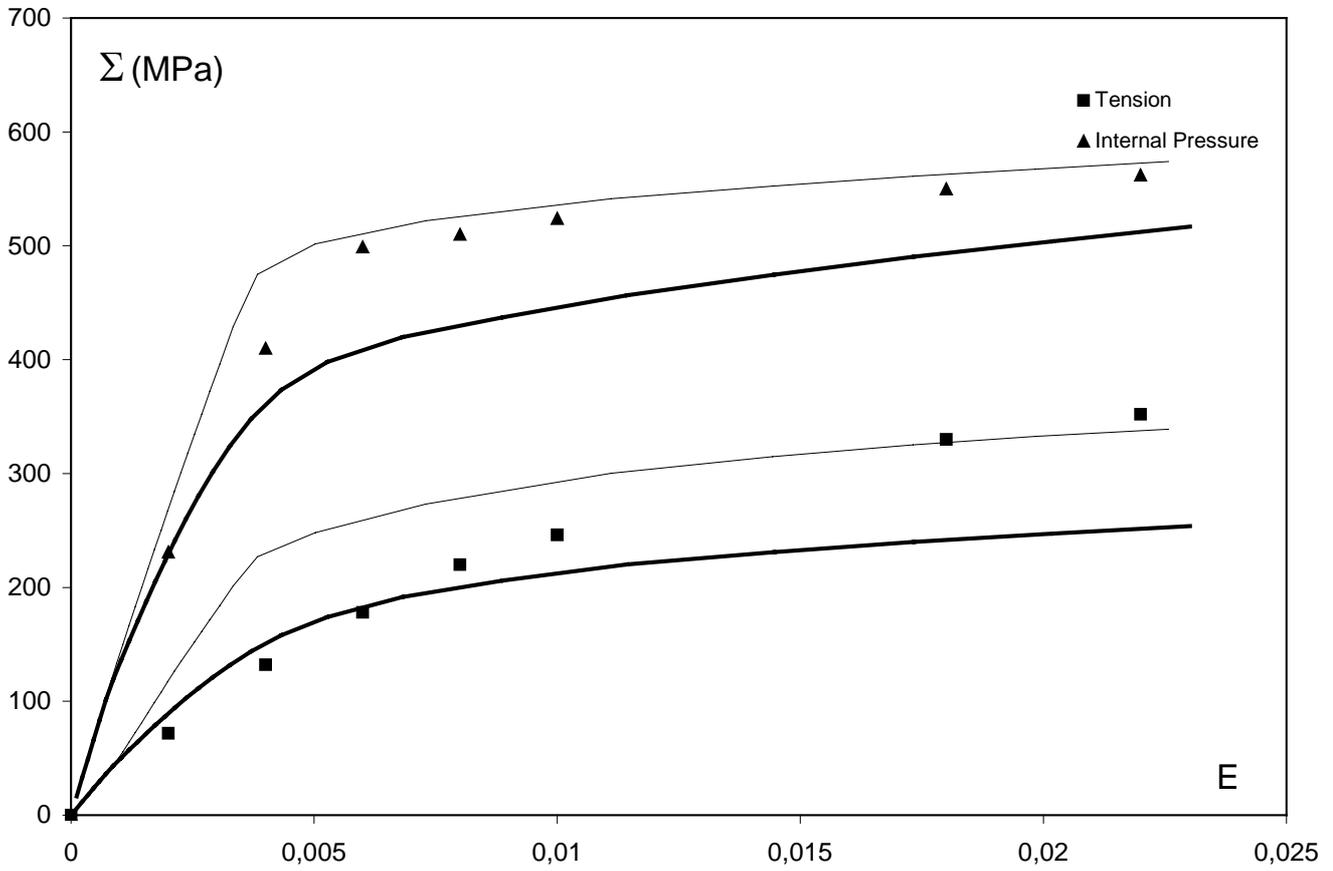

Figure 15